\documentclass{article}

\usepackage{arxiv}
\usepackage[utf8]{inputenc} % allow utf-8 input
\usepackage[T1]{fontenc}    % use 8-bit T1 fonts
\usepackage{hyperref}       % hyperlinks
\usepackage{url}            % simple URL typesetting
\usepackage{booktabs}       % professional-quality tables
\usepackage{amsfonts}       % blackboard math symbols
\usepackage{nicefrac}       % compact symbols for 1/2, etc.
\usepackage{microtype}      % microtypography
\usepackage{lipsum}		% Can be removed after putting your text content
\usepackage{graphicx}
\usepackage{doi}
\usepackage{verbatim}
\usepackage{authblk}
\usepackage{xcolor}
\usepackage{fancyvrb}
\usepackage{fvextra}
\usepackage{subcaption} 
\usepackage{caption}
\title{DT4PCP: A Digital Twin Framework for Personalized Care Planning Applied to Type 2 Diabetes Management}

\author{%
    \textbf{Javad M Alizadeh}\textsuperscript{1,\href{https://orcid.org/0000-0003-1421-8281}{\includegraphics[scale=0.06]{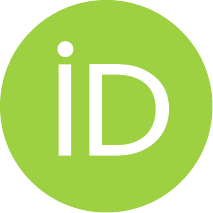}}},
    \textbf{Mukesh K Patel}\textsuperscript{2,\href{https://orcid.org/0009-0004-4456-2870}{\includegraphics[scale=0.06]{orcid.pdf}}}, 
    \textbf{Huanmei Wu}\textsuperscript{3,\href{https://orcid.org/0000-0003-0346-6044}{\includegraphics[scale=0.06]{orcid.pdf}}}, 
   }

\affil{%
    \textsuperscript{1}{ javad.mohammad.alizadeh@temple.edu} \\
    \textsuperscript{2}{ mukesh.kumar.patel@temple.edu} \\
    \textsuperscript{3}{ huanmei.wu@temple.edu} \\
    \textsuperscript{1,2,3}Department of Health Services Administration and Policy, College of Public Health, Temple University, Philadelphia, Pennsylvania, USA
}

\begin{document}
\maketitle

\begin{abstract}
	Digital Twin (DT) technology has emerged as a transformative approach in healthcare, but its application in personalized patient care remains limited. This paper aims to present a practical implementation of DT in the management of chronic diseases. We introduce a general DT framework for personalized care planning (DT4PCP), with the core components being a real-time virtual representation of a patient’s health and emerging predictive models to enable adaptive, personalized care. We implemented the DT4PCP framework for managing Type 2 Diabetes (DT4PCP-T2D), enabling real-time collection of behavioral data from patients with T2D, predicting emergency department (ED) risks, simulating the effects of different interventions, and personalizing care strategies to reduce ED visits. The DT4PCP-T2D also integrates social determinants of health (SDoH) and other contextual data, offering a comprehensive view of the patient's health to ensure that care recommendations are tailored to individual needs. Through retrospective simulations, we demonstrate that integrating DTs in T2D management can lead to significant advancements in personalized medicine. This study underscores the potential of DT technology to revolutionize chronic disease care.
\end{abstract}

% keywords can be removed

\section{Introduction}
T2D is a chronic condition and a major global health challenge, affecting millions of individuals and placing a substantial burden on healthcare systems~\cite{gallardo2021type, quinones2021geographically, lawrence2021trends}. When poorly controlled, T2D can lead to severe complications such as hyperglycemia, cardiovascular disease, kidney failure, and neuropathy, often resulting in ED visits~\cite{Alizadeh2024, lindekilde2022prevalence}. Traditional management strategies rely on periodic clinical visits and treatment plans based on general T2D guidelines~\cite{samson2023american}, which may not fully address individual patient variability or account for real-time health changes. With the rise of digital health technologies, there is an increasing demand for personalized, data-driven solutions that can predict risks and proactively guide interventions to prevent severe complications before they necessitate emergency care~\cite{vuohijoki2020implementation,mikkola2020personalized,mikkola2022association}.

DT technology provides a groundbreaking approach to addressing healthcare challenges~\cite{Katsoulakis2024}. A DT is a real-time virtual representation of an object that integrates both retrospective and dynamic data sources. A human digital twin for health represents a person’s health conditions, including demographics, SDoH, clinical records, and lifestyle behaviors, to enhance predictions and support timely medical decisions~\cite{Zhang2024}. By continuously learning from patient data, DT models can simulate disease progression, predict emergency risks, and recommend targeted interventions tailored to individual needs~\cite{Vallee2024}. This shift enables healthcare providers to transition from reactive treatment to proactive disease management, thereby reducing avoidable ED visits and improving long-term patient outcomes.

Recent studies show that DT-driven treatment can improve blood sugar control, lower HbA1c levels, and promote adherence to treatment plans, 
%all contributing to 
resulting in fewer diabetes-related complications~\cite{Shamanna2024}. Machine learning approaches enable DTs to dynamically adjust as a patient’s condition evolves, ensuring personalized recommendations that align with real-time health data~\cite{Rad2024a}. Additionally, scenario-based simulations within DT frameworks empower clinicians to evaluate different intervention strategies before implementing them, improving treatment efficacy and patient safety~\cite{Cellina2023}. 

This paper presents a practical implementation of DT technology in managing chronic diseases, specifically T2D. We will first introduce the DT for Personalized Care Planning (\textit{i.e.}, DT4PCP) framework for chronic health conditions, which integrates real-time health data with emerging predictive models to enable adaptive, customized care. Next, we will demonstrate how the framework can be applied to T2D management (\textit{i.e.}, DT4PCP-T2D) to predict emergency department (ED) risks and provide care strategies with the goal of reducing ED visits~\cite{Shamanna2020, Katsoulakis2024}.

\section{The DT4PCP Framework }
\begin{figure*}[t] % Use figure* for full-page-width figures
  \centering
  \includegraphics[width=\textwidth]{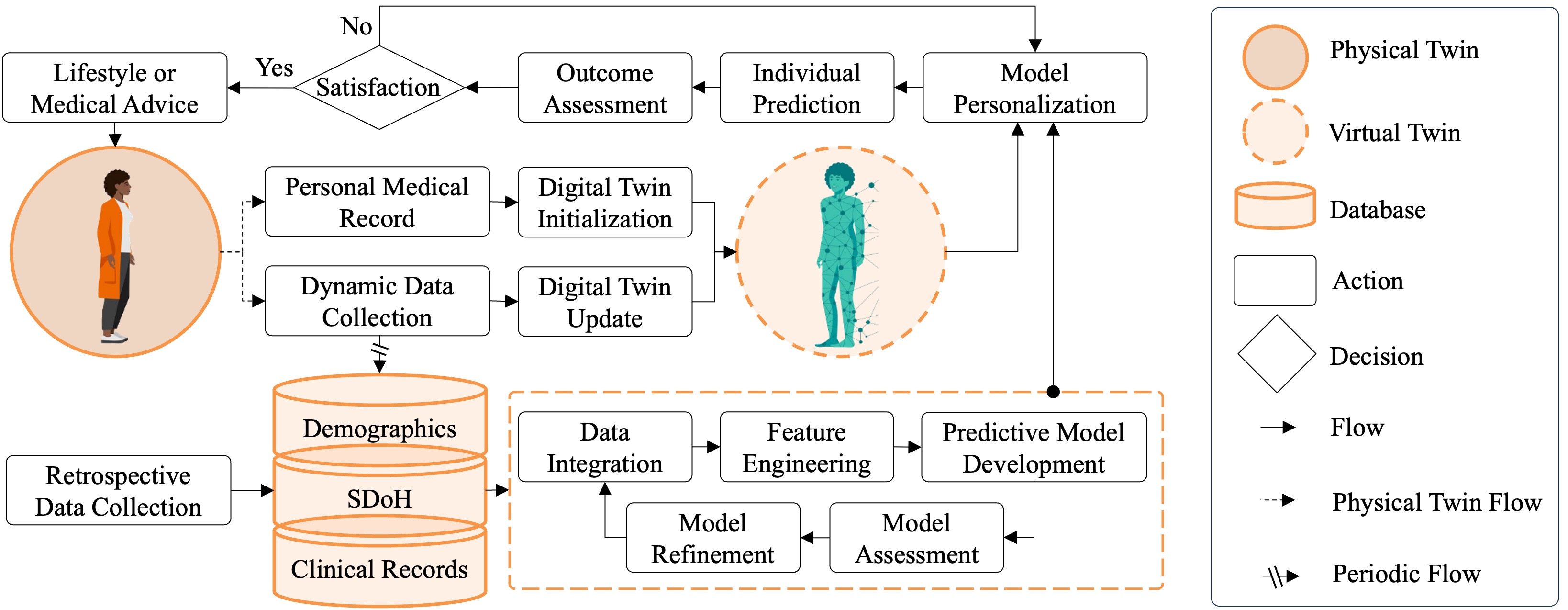} % Replace "test" with your image filename
  \caption{Digital Twin Framework.}
  \label{fig:fw} % Add a label for cross-referencing
\end{figure*}

As illustrated in Figure~\ref{fig:fw}, the proposed DT for personalized care planning (DT4PCP) offers a dynamic framework for integrating DT technology into healthcare. The DT4PCP framework combines trained machine learning (ML) models with real-time patient health data to predict emergency department (ED) visits. It also simulates the effects of various interventions with different parameters to help generate personalized recommendations. To facilitate the use of the DT4PCP, we have developed a user-friendly Graphical User Interface (GUI) that allows healthcare providers and caregivers to interact with the system seamlessly. At the core of this framework is the creation of a comprehensive, real-time virtual representation of the health conditions of a patient, enabling adaptive care adjustments and optimized treatment strategies for improved health outcomes.

\subsection{The DT Source Data}
The source data for the DT4PCP framework consists of two key components: retrospective data collection and dynamic data collection.
The retrospective data collection includes a range of multimodal data sources, such as electronic health records (EHR) for clinical histories and chronic disease information, social determinants of health (SDoH), demographics, and aggregated personal behavioral data (e.g., smoking, alcohol consumption, and other lifestyle factors). This data is used to train the machine learning (ML) models. The more comprehensive and extensive the retrospective data, the better the performance of the trained predictive models, and consequently, the DT4PCP system. High-quality data enables the generation of more accurate models, improving the system's overall effectiveness. Additionally, this data can be periodically updated, allowing for the continuous refinement of pretrained models, which further enhances the system performance over time.

The dynamic data, including vital signs, physical activity, and clinical visits, are continuously collected in real-time, enabling the DT4PCP to function effectively for monitoring, prevention, and treatment of specific chronic conditions. This real-time data feeds into the digital patient (\textit{i.e.}, the individual human digital twin), providing a comprehensive and up-to-date representation of the patient’s health status. With this data, the system can generate predictions based on expected health outcomes. The continuous flow of dynamic data ensures ongoing monitoring of the patient’s health, allowing for timely interventions and updates to care strategies as the patient's condition evolves.

\subsection{Generate the Pretrained Models}
The DT4PCP leverages various ML models to predict different health outcomes based on relevant health data. 
There are five key steps to generate the pretrained models based on retrospective data collection: 1) data integration, 2) feature engineering, 3) predictive model development, 4) model assessment, and 5) model refinement.
The pretrained model will be retrained when sufficient new data becomes available, when new algorithms are introduced, or when the system's performance requires improvement.

For the sample DT4PCP-T2D implementation, the primary goal is to use ML models to predict near-future emergency department (ED) visits for patients with Type 2 Diabetes (T2D). This enables healthcare professionals and caregivers to adopt targeted intervention strategies or preventative measures to mitigate the risk of ED visits.
For model development, algorithms such as CatBoost (CatB), Ensemble Learning (EL), K-Nearest Neighbors (KNN), Support Vector Classification (SVC), Random Forest (RF), and eXtreme Gradient Boosting (XGB) were utilized, and they were optimized through cross-validation. The models were then evaluated using metrics like accuracy, precision, and AUC to identify the best-performing predictive model.

\subsection{Personalized Individual DT4PCP}
The DT4PCP for an individual is initialized when their personal medical records become available. The individual’s DT is then continuously updated using dynamic data automatically collected from clinical visits or manually entered by verified personnel.
The collected real-time data plays a crucial role in enhancing the quality of future predictions With the pretrained models, the ML models are personalized based on the specific individual's annotated data. 
Personalized recommendations are generated by healthcare providers. 
If the outcomes are satisfactory, the patient receives provider-generated lifestyle or medical recommendations. If not, healthcare providers adjust the recommendations, and the personalized ML models learn from these updated inputs using augmented methods to refine the models. This iterative process continues until the provider is confident in the outcome.
The real-time data and the final recommendations are stored for future ML model personalization. This personalized ML model training occurs periodically, with the frequency determined by the administrator based on the volume of stored data. 
 
\section{Apply DT4PCP to T2D Management}
We implemented the DT4PCP framework for reducing ED visits among patients with T2D. The machine learning (ML) models were trained using data extracted from the HealthShare Exchange (HSX) Clinical Data Repository, which included 34,151 adults with T2D who had at least one clinical encounter at participating HSX hospitals between January 1, 2017, and December 31, 2022.

To enhance data validity, we excluded patients with hypertension, children with T2D, and those diagnosed with Type 1 Diabetes. The dataset encompassed over 76 million medical encounters, 113 million vital sign records, and 123 million diagnoses. Social determinants of health (SDoH) factors from the American Community Survey and SEDAC were linked by zip code to provide a broader contextual view. Data preprocessing ensured consistency by standardizing vital sign units (e.g., height in cm/inches, weight in pounds/kilograms) and harmonizing diagnostic codes across ICD-9, ICD-10, and SNOMED-CT, reducing 65,000 unique diagnoses to 742 categories.

Finally, feature selection focused on demographics, diagnoses, SDoH indicators, and vital signs. The number of diagnosis features was reduced from 2,498 to the 30 most frequent ones to mitigate overfitting.

\subsection{Performance of Pretrained Models}
Pretrained models were generated using CatBoost (CatB), Ensemble Learning (EL), K-Nearest Neighbors (KNN), Support Vector Classification (SVC), Random Forest (RF), and eXtreme Gradient Boosting (XGB). To predict ED visits, machine learning (ML) models were trained on a 70\% training and 30\% testing split, using 10-fold cross-validation. The outcome was classified as binary, where 0 represented no ED visit and 1 indicated an ED visit. Model performance was evaluated using accuracy, precision, recall, F1-score, and Area Under the Curve (AUC).

\begin{figure}[h]
  \centering
  \begin{minipage}[b]{0.48\textwidth}
    \includegraphics[width=\linewidth]{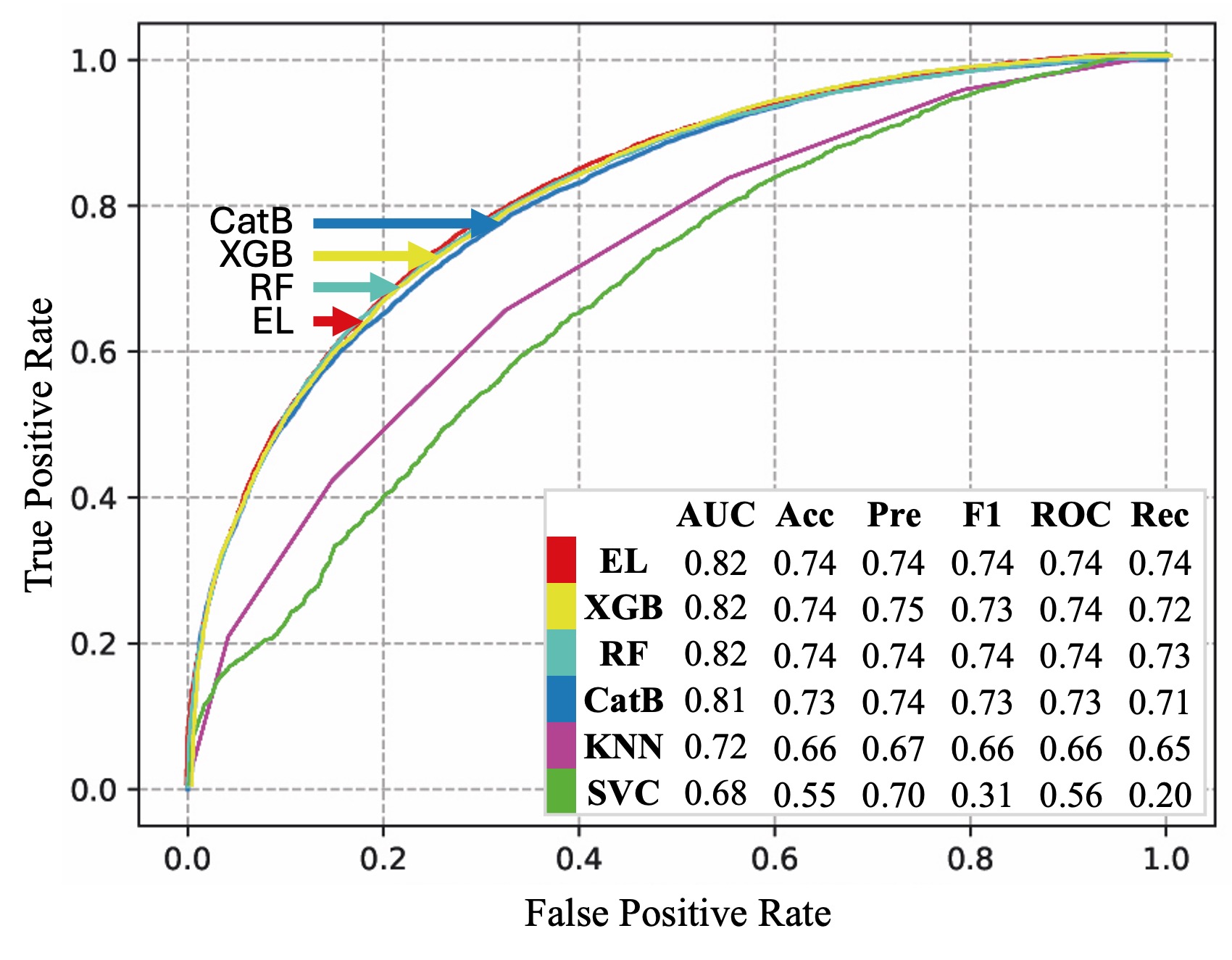}
    \captionof{figure}{The Receiver Operator Characteristic (ROC) curves of the predictive models and their corresponding evaluation metrics.}
    \label{fig:metrics}
  \end{minipage}
  \hfill
  \begin{minipage}[b]{0.48\textwidth}
    \includegraphics[width=\linewidth]{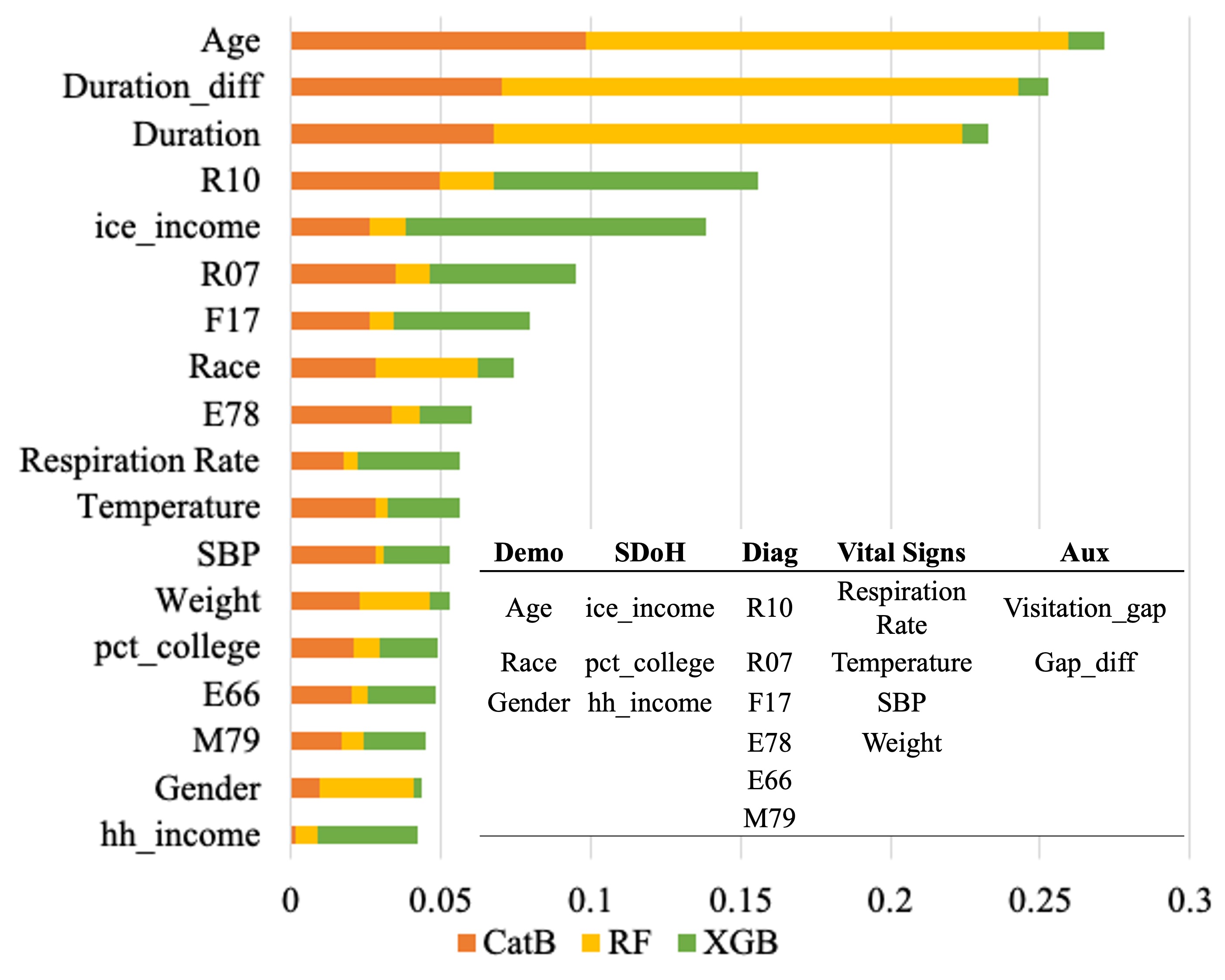}
    \captionof{figure}{Features and the feature importance obtained from the three ML models.}
    \label{fig:featuresIm}
  \end{minipage}
\end{figure}

As shown in Figure \ref{fig:metrics}, EL, XGB, and RF achieved the highest AUC of 0.82, with EL and RF showing balanced metrics (accuracy, precision, recall, and F1-score around 0.74). XGB had slightly higher precision (0.75) but lower recall (0.72). CatB followed closely with an AUC of 0.81 and metrics around 0.73, remaining competitive. KNN underperformed with an AUC of 0.72 and metrics around 0.66–0.67, suggesting limited suitability. SVC produced the weakest results, with an AUC of 0.68, accuracy of 0.55, and recall of 0.20, frequently missing positive cases.

As shown in Figure \ref{fig:featuresIm}, feature importance analysis using CatB, RF, and XGB identified key predictors across demographics, SDoH, diagnoses, vital signs, and healthcare utilization. Demographic factors, including age, race, and gender, were found to influence ED visits, with older patients at higher risk. SDoH factors such as income, education, and household income highlighted the socioeconomic impact on healthcare access. Diagnoses, including ICD-10 codes for abdominal pain (R10), chest pain (R07), nicotine dependence (F17), and obesity (E66), were strongly associated with T2D complications. Vital signs, such as respiration rate, temperature, systolic blood pressure (SBP), and weight, played a critical role in risk assessment. Additionally, healthcare utilization patterns, such as gaps in visit frequency, were important predictors of ED visits. 

\subsection{Graphical User Interface}

\begin{figure*}[t] % Use figure* for full-page-width figures
  \centering
  \includegraphics[width=\textwidth]{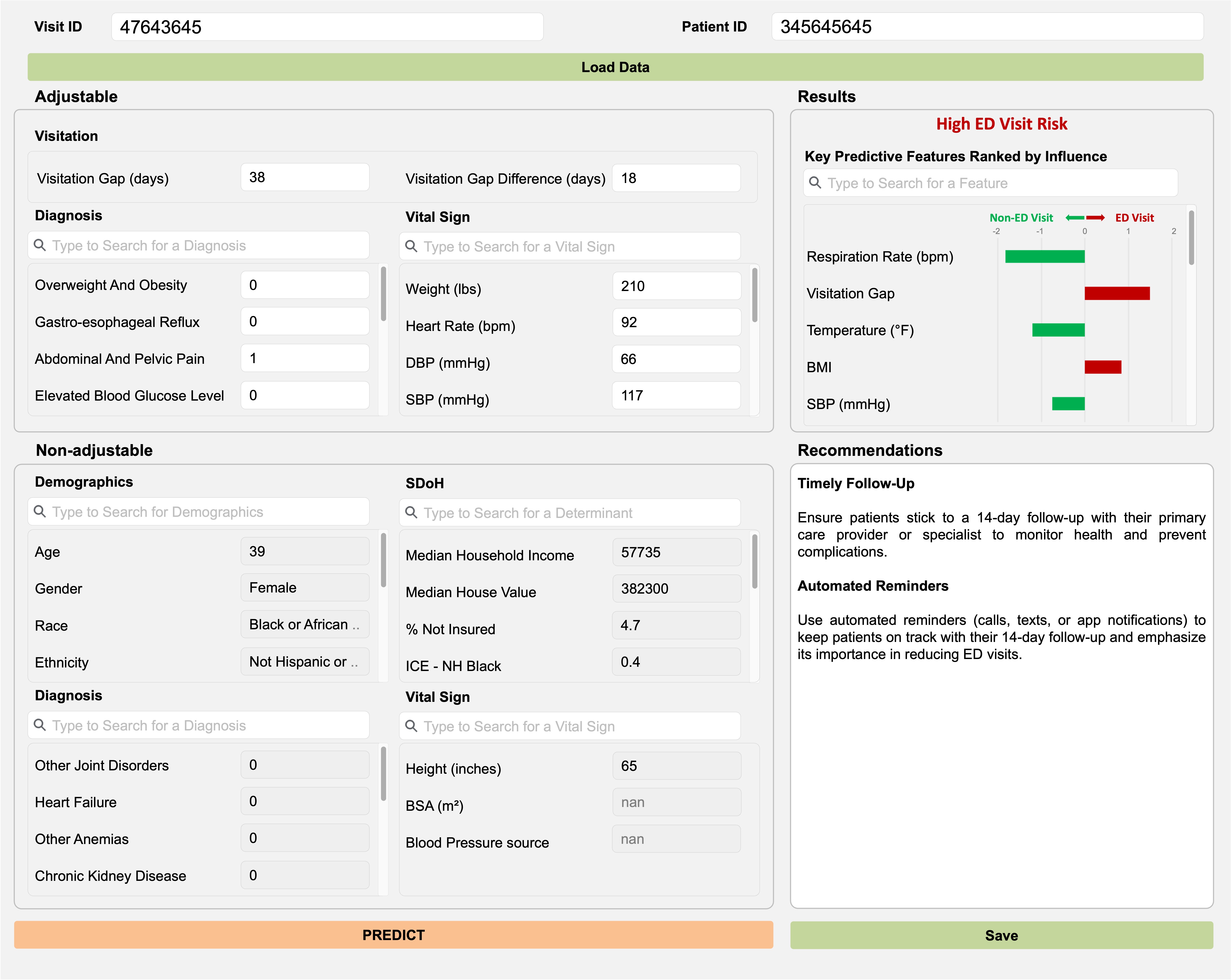} % Replace "test" with your image filename
  \caption{An illustration of the graphical user interface demonstrating its main features.}
  \label{fig:gui} % Add a label for cross-referencing
\end{figure*}

A graphical user interface (GUI) was designed to allow healthcare providers to use the DT4PCP system effectively. 
As shown in Figure \ref{fig:gui}, providers enter the patient and visit ID, prompting the system to retrieve personal medical records and relevant dynamic data from the EHR systems. After loading the data, the interface displays adjustable features, such as diagnoses and vital signs, and non-adjustable features, such as demographics and SDoH, along with their corresponding values and units in designated fields. Providers can search for additional features and modify the values of adjustable features if necessary. By clicking the "PREDICT" button, the system generates prediction results. These results indicate whether a patient has a high or low risk of an ED visit and include key predictive features ranked by their SHAP values. Providers can further modify feature values based on their influence on the outcome and re-run the prediction as needed until they are satisfied with the results. Finally, based on the feature influences of adjustable parameters and their normal ranges for the specific patient, the provider offers personalized recommendations to help reduce the risk of future ED visits. These recommendations are entered into a free-text box, which providers can edit and save for further clinical action.

To illustrate the process, consider the following example scenario to use the DT4PCP-T2D by providers with the following 11 steps.

\vspace{3pt}
\noindent \textbf{DT4PCP-T2D Initiation}:
\vspace{-3pt}
    \begin{enumerate}
        \item The provider enters the patient ID and visit ID into the GUI.
        \item The system retrieves the patient's data from the EHR.
        \item The GUI lists adjustable and non-adjustable features.
    \end{enumerate}

\noindent \textbf{Provider input}:
\vspace{-3pt}
    \begin{enumerate}  \setcounter{enumi}{3}
        \item The provider modifies the values of adjustable features (e.g., BMI, SBP) as needed in the GUI.
        \item The DT is automatically updated based on the entered data.
    \end{enumerate}
    
\noindent \textbf{Prediction}:
\vspace{-3pt}
    \begin{enumerate}  \setcounter{enumi}{5}
        \item The provider clicks the "PREDICT" button, and the system generates prediction results, including the risk of an ED visit and predictive features ranked by their influence.
        \item The system identifies key risk factors for ED visits (e.g., high visitation gap of 38 days).
    \end{enumerate}
    
\noindent \textbf{Re-adjustment and Re-prediction}:
\vspace{-3pt}
    \begin{enumerate}  \setcounter{enumi}{7}
        \item  The provider can modify key feature values (e.g., BMI, SBP) in the GUI and click the "PREDICT" button again to update the results.
        \item  The provider reviews the results and can modify other adjustable features for re-prediction as needed.
    \end{enumerate}
    
\noindent \textbf{Personalized Recommendations}:
\vspace{-3pt}
    \begin{enumerate}  \setcounter{enumi}{9}
        \item  Based on the updated results, the provider offers personalized recommendations (e.g., weight-loss program, dietary changes, antihypertensive therapy).
        \item  The provider enters the recommendations and saves them for future monitoring and updates.
    \end{enumerate}

\section{Discussion}

The implementation of a DT framework for predicting ED visits in T2D patients offers a transformative approach to personalized and data-driven healthcare. By integrating real-time patient data with ML algorithms, this system addresses key challenges in T2D management, including predicting ED visits, targeting modifiable risk factors, and improving clinical decision-making. The following discussion highlights the system's strengths, potential applications, limitations, and implications for future research and practice.

This framework provides real-time, personalized insights by integrating comprehensive patient data, including demographics, environment, healthcare utilization, and health status. By accounting for both clinical and non-clinical factors influencing ED visits, it enables precise, actionable recommendations that help providers intervene early and reduce adverse outcomes. A continuous feedback loop ensures the system adapts as patient health evolves, making it particularly effective for managing chronic conditions like T2D. Providers can use the GUI to simulate intervention outcomes before real-world implementation, enhancing evidence-based decision-making and reducing uncertainty.

Beyond individual patient care, the system benefits healthcare providers, administrators, and policymakers. It identifies high-risk patients, highlights key predictors of ED visits, and supports targeted resource allocation. Patients gain proactive engagement through personalized recommendations, while early risk warnings help prevent complications. On a broader scale, aggregated insights inform public health strategies, addressing disparities and improving health equity. By reducing unnecessary ED visits, the system enhances efficiency and supports cost containment efforts in healthcare.

While this framework demonstrates significant potential, several limitations must be addressed. Expanding its application to patients with both T2D and hypertension could enhance its ability to manage comorbid conditions. Incorporating real-time data from wearable devices may improve predictive accuracy and system responsiveness. Additionally, integrating advanced technologies such as blockchain for secure data sharing could further strengthen its capabilities. Future research should assess the long-term impact on patient outcomes and healthcare costs by measuring reductions in ED visits, improved glycemic control, and patient satisfaction.

Furthermore, automating the simulation phase to generate meaningful visit scenarios for individual patients enhances the usability and robustness of the framework. By allowing features to fluctuate within their ranges, critical thresholds can be identified. These thresholds represent turning points that significantly influence predictions. This process can be achieved through advanced computational methods such as optimization algorithms. These algorithms systematically explore feature variations to quantify their impact on outcomes.

Additionally, integrating large language models (LLMs) can improve the framework's capability and usability for healthcare providers. LLMs can generate preliminary recommendations based on prediction results when combined with strategic prompt engineering. Such integration streamlines clinical decision-making and enhances the interpretability and accessibility of the framework. This makes the framework a more practical and actionable tool in real-world clinical settings.

From a technical perspective, while multiple ML algorithms were evaluated, the study does not explore deep learning models, which could improve predictive performance, particularly with large, complex datasets. Moreover, periodic model retraining is dependent on data volume, potentially delaying updates and reducing adaptability to sudden health changes. The implementation of this framework in clinical settings presents ethical and practical challenges. Ensuring data privacy and security is crucial, necessitating clear governance protocols for patient consent and secure storage. Effective provider training is also essential for seamless integration into existing workflows. The system’s success depends on user-friendly interfaces and ongoing support to ensure healthcare providers can interpret and act on recommendations effectively. Addressing these challenges will be key to optimizing the framework’s real-world impact.

\section{Conclusion}

This study developed a dynamic digital twin framework for personalized care planning (DT4PCP) and have a prototype implementation to enhance personalized T2D management, predict ED visits by patients with T2D, and reduce complications through proactive interventions. The framework, featuring Ensemble Learning, Random Forest, and XGBoost models, achieved strong predictive performance with an AUC of 0.82. Key predictors of ED visits included SDoH, clinical factors (e.g., obesity, blood pressure), and healthcare utilization patterns. Iterative model refinement and feature engineering enabled patient-specific risk assessments. \\
The framework supports clinicians by identifying high-risk patients, simulating interventions, and tailoring recommendations, while also helping policymakers optimize resource allocation and address health disparities. By incorporating SDoH and contextual data, it provides a comprehensive view of patient health, ensuring personalized care and improving patient outcomes. This approach integrates predictive analytics with real-world data to enhance decision-making, patient engagement, and diabetes management, ultimately reducing avoidable ED visits and healthcare costs.

\section{Acknowledgments}
The project is in part supported by the Robert Wood Johnson Foundation’s Health Data for Action program, managed by AcademyHealth (Grant\# 78961).

%% the bibliography file.
\bibliography{References}

\end{document}